\documentclass[reprint,amsmath,amssymb,aps,prl]{revtex4-1}
\usepackage{graphicx} 
\usepackage{epstopdf}
\usepackage{amssymb}
\usepackage{dcolumn} 
\usepackage{bm} 
\usepackage{ulem} 
\usepackage{amsmath}
\usepackage{color}
\usepackage{threeparttable}
\usepackage{graphicx,rotating,booktabs}
\usepackage{comment}
\usepackage[colorlinks,allcolors=blue]{hyperref}

\begin{document}
\title{Nondestructive Prediction of the Buckling Load of Imperfect Shells} 

\author{Ana\"{i}s Abramian$^{1,2}$}
\author{Emmanuel Virot$^{2,3}$}
\author{Emilio Lozano$^{2,3}$}
\author{Shmuel M. Rubinstein$^{2}$}
\author{Tobias M. Schneider$^{3}$}
\affiliation{$^1$Institut Jean le Rond d'Alembert, Sorbonne Universit\'{e}, CNRS, UMR 7190, 75005 Paris, France}
\affiliation{$^2$John A, Paulson School of Engineering and Applied Sciences, Harvard University, Cambridge, MA 02138, USA}
\affiliation{$^3$Emergent Complexity in Physical Systems Laboratory (ECPS), \'Ecole Polytechnique F\'ed\'erale de Lausanne, CH 1015 Lausanne, Switzerland}

\date{Received 17 May 2020; revised 7 July 2020; accepted 13 October 2020; published 25 November 2020}

\begin{abstract}

From soda cans to space rockets, thin-walled cylindrical shells are abundant, offering exceptional load carrying capacity at relatively low weight. However, the actual load at which any shell buckles and collapses is very sensitive to imperceptible defects and can not be predicted, which challenges the reliable design of such structures. Consequently, probabilistic descriptions in terms of empirical design rules are used and reliable design requires to be very conservative. We introduce a nonlinear description where finite-amplitude perturbations trigger buckling. Drawing from the analogy between imperfect shells which buckle and imperfect pipe flow which becomes turbulent, we experimentally show that lateral probing of cylindrical shells reveals their strength non-destructively. A new ridge-tracking method is applied to commercial cylinders with a hole showing that when the location where buckling is nucleated is known we can accurately predict the buckling load of each individual shell, within $\pm 5\%$. Our study provides a new promising framework to understand shell buckling, and more generally, imperfection-sensitive instabilities.\newline

\noindent DOI: \href{https://journals.aps.org/prl/abstract/10.1103/PhysRevLett.125.225504}{10.1103/PhysRevLett.125.225504}

\end{abstract}

\maketitle 

Compress an empty soda can from its top and it will remain stable over a considerable range of loads. 
However, at a critical load, the can eventually buckles, violently collapses, and irreversibly deforms.  
A classical linear stability analysis greatly overestimates the buckling load of thin cylinder shells and domes \cite{Seide1960, Yamaki1984, NASA_SP8032}, because the system is extremely sensitive to defects \cite{Koiter1945,Weingarten1968,Horton1965,Bushnell981,Davis1987,Singer2002}. 

\begin{figure}[b]
\centering
\includegraphics[width = 0.9\columnwidth]{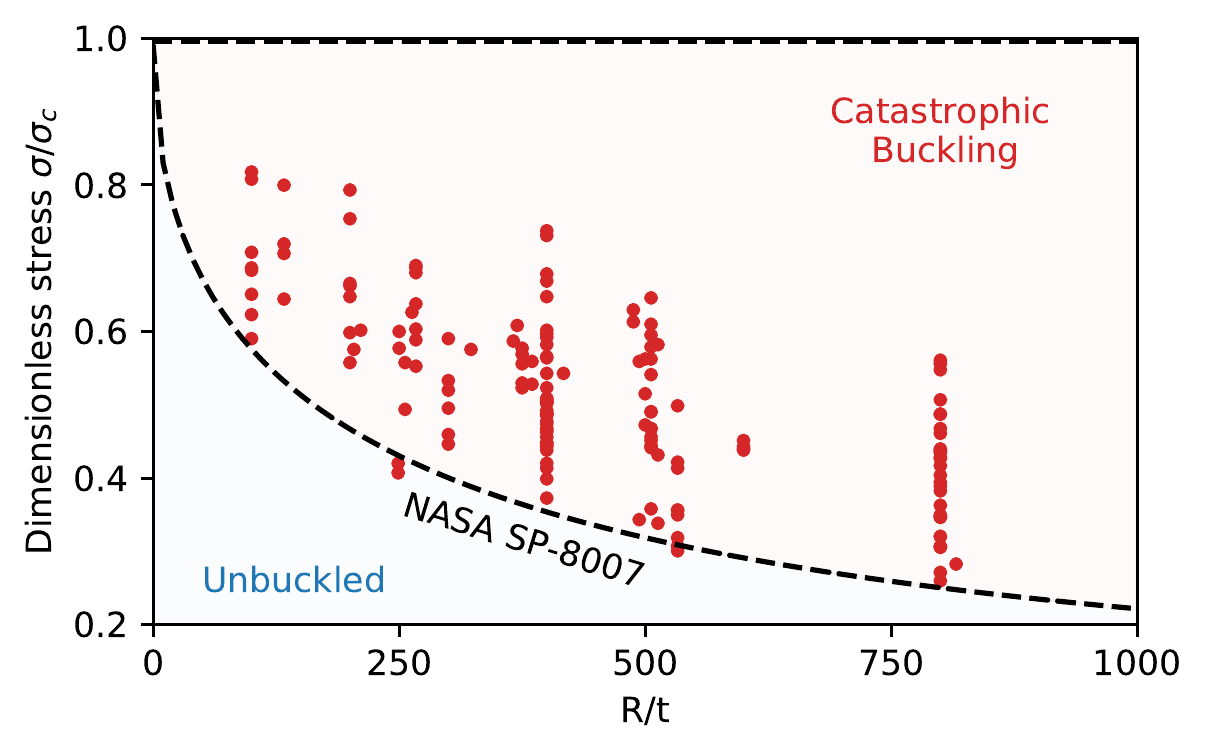}
\caption{
Knockdown factor $\sigma/\sigma_c$ vs $R/t$ collected by NASA with the empirical design rule indicated by the dashed line. (Data from Seide et al. (1960) \cite{Seide1960})
}
\label{fig:nasa}
\end{figure}

Classical theory mostly cannot capture the influence of defects and thus, in general dramatically overestimates the carrying-load of engineered shells, infamously affecting structural reliability \cite{Hilburger2015}. 
The ratio between the actual measured buckling load and the theoretical prediction is termed the \textit{knock-down factor}. The knock-down factor tends to decrease with the ratio of the shell's radius of curvature to its thickness, $R/t$, but there are extreme stochastic variations between nominally identical shells, as shown in Fig.~\ref{fig:nasa}.
Fifty years ago, NASA SP-8007 proposed a conservative phenomenological ``design rule" \cite{Weingarten1968}, which estimates a lower bound for the distribution of knock-down factors and remains the basis for US and European design codes even today. 

Any sound prediction of the knock-down factor of a specific manufactured shell currently requires complete knowledge of the shell's imperfections. A full characterization of imperfections is extremely challenging \cite{Hilburger2015} and practically impossible for real engineered structures. Consequently, the NASA design rule which on average severely underestimates buckling-loads remains still in use. Here we introduce a non-destructive method that accurately predicts the buckling load of an individual defected shell without prior knowledge of imperfections.

Characterizing defect-sensitivity as observed in shell buckling is notoriously difficult because in any experimental setup a multitude of details need to be controlled practically perfectly. 
Notably, a similarly challenging situation arose in fluid mechanics, where the concepts of finite-amplitude perturbations in combination with very careful experiments lead to a complete rethinking of the transition from laminar flow to turbulence in pipe flows \cite{Darbyshire1995,Grossmann2000,Kerswell2005,Eckhardt2007}. 
Just as there is no universal buckling load at which real imperfect shells buckle, there is no universal Reynolds number at which turbulence develops in real, imperfect pipes; and the critical flow-speed at which turbulence occurs is highly sensitive to imperfections. 

\begin{figure}
\centering
\includegraphics[width =\columnwidth]{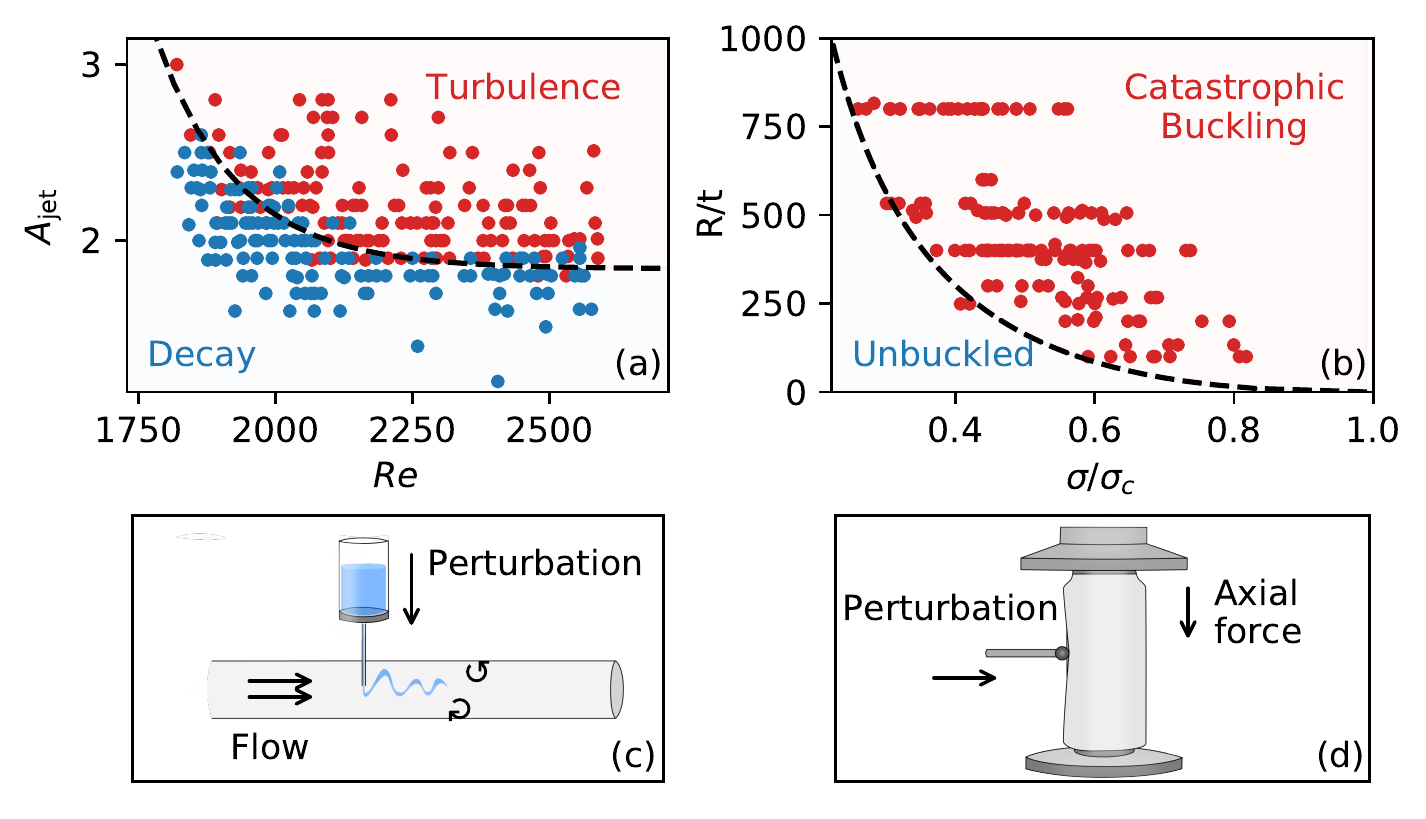}
\caption{
(a) $A_\text{jet}$ vs Re for experiments by Darbyshire $\&$ Mullin \cite{Darbyshire1995} (b) $R/t$ vs knockdown factor $\sigma/\sigma_c$ for Nasa’s data (Fig. \ref{fig:nasa}). 
(c/d) Lateral poking of the shell corresponds to the transverse jet perturbation and probes the stability of the linearly stable shell to well-controlled perturbations. 
}
\label{fig:analogy}
\end{figure}

The challenging problem of the transition to turbulence is tackled within a nonlinear framework by considering the response of a perfect system, an ideal pipe, to a jet perturbation of finite-amplitude, $A_\text{jet}$, that can trigger turbulence \cite{Darbyshire1995}. 
In the context of shell buckling, an analogous approach implies revisiting the NASA data, shown in Fig.~\ref{fig:nasa}, by considering the axial load as an externally-imposed control parameter, like the Reynolds number. 
Since controlling manufacturing imperfections is harder for thinner shells, the geometrical quantity $R/t$ (radius over thickness) should be treated as a proxy for the vulnerability to defects of a given size. With this interpretation, the phenomenology of the transition to turbulence in pipe flow and the buckling of cylindrical shells appears strikingly similar, as shown in Fig.~\ref{fig:analogy}(a,b).

Inspired by the hydrodynamic analogy, we recently developed an experimental and theoretical framework \cite{Virot2017} to study shell buckling. The jets, which trigger turbulence in pipes correspond to a lateral probe, or a ``poker'', in our system, as shown on Fig.~\ref{fig:analogy}(c-d). 
Similar ideas were recently pursued by Hutchinson, Thompson and coworkers \cite{Thompson2015,Thompson2016,Hutchinson2017a,Hutchinson2017b} as well as others \cite{Neville2018,Groh2019a,Groh2019b}. Analogous experiments were performed by Reis and coworkers for pressurized hemispherical shells \cite{Lee2016,Marthelot2017} and cylindrical shells with radially applied perturbation loads were studied in the context of aerospace applications \cite{Craig1973,Wullschleger2006,Huhne2008,Kriegesmann2012,Haynie2012,Arbelo2014,Hao2015,Wagner2018,Jiao2018}. 


In contrast to previous work, in our approach the poker-induced deformation is not interpreted as a defect; instead, the poker is a probe exploring the system's nonlinear response to finite-amplitude perturbations. By measuring the poker force applied on the shell $F_P$ as a function of displacement, $D_P$, we recently quantified the critical perturbations that triggers buckling for varying axial loads $F_A$. 
Mapping out the conditions under which buckling occurs, we identified a universal stability landscape for cylindrical shells, represented as a surface in a ($F_A$, $F_P$, $D_P$) three-dimensional space \cite{Gerasimidis2018, Virot2017}. 

\begin{figure}[h]
\centering
\includegraphics[width =\columnwidth]{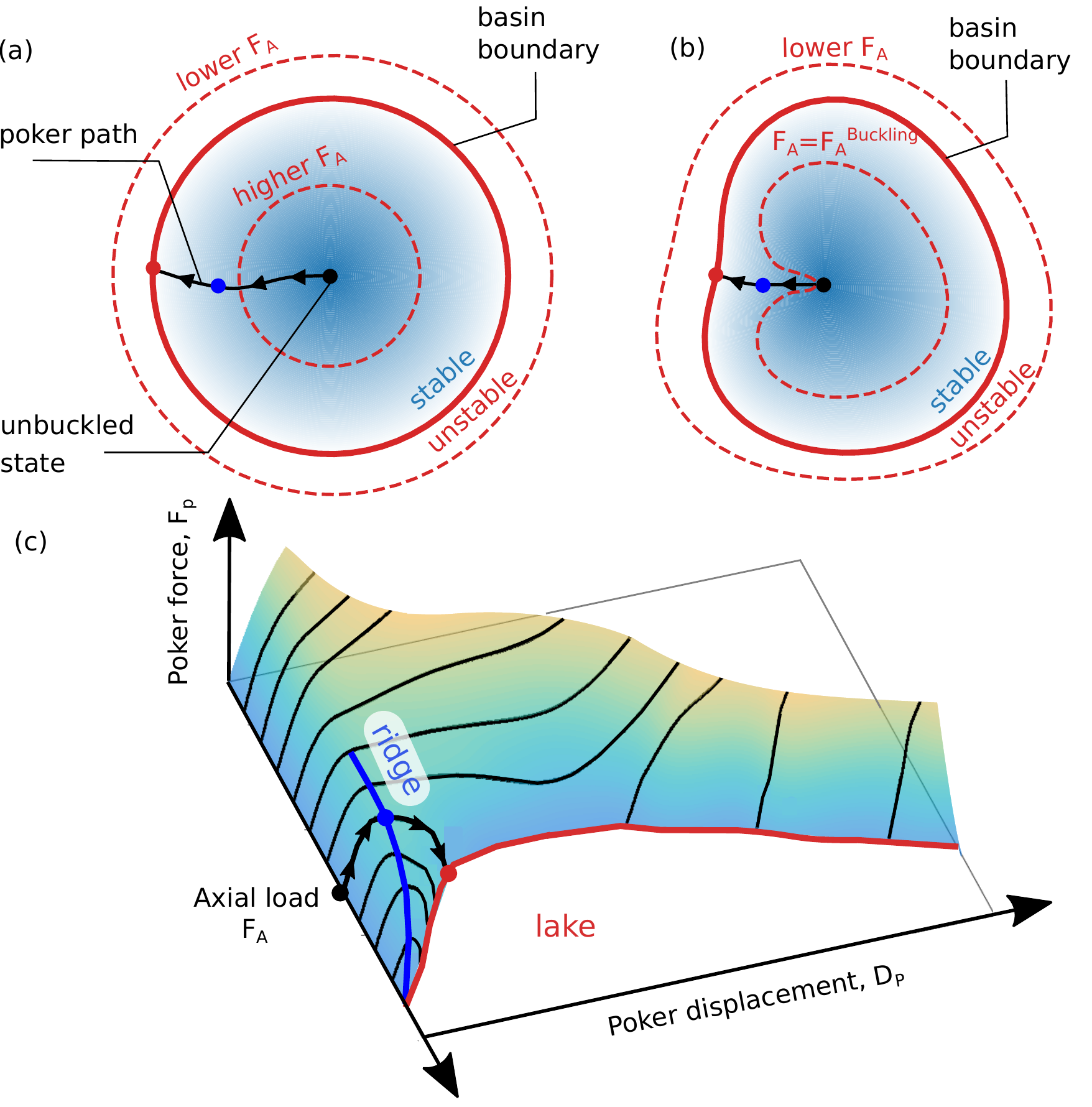}
\caption{
Schematic 2D projection of the system's state-space for a perfect shell (a) and one with imperfections (b). 
The stability landscape (c) probes the basin boundary along a path of dimple-shaped deformations induced by lateral poking. The point on the ridge (blue dot) is located between the basin boundary and the unbuckled state. It varies with $F_A$ creating the ridge (blue line in (c)) that asymptotes to zero when the basin of attraction vanishes and spontaneous buckling is induced. Shell imperfections distort the basin boundary and modify its experimentally accessible representation, the stability landscape. Ridge-tracking reveals the influence of imperfections while non-destructively probing the ridge. 
}
\label{fig:stability_landscapes}
\end{figure}

The landscape, which can be probed experimentally, characterizes the stability of a shell, which, in the most general case, is dictated by the complex correlation structure of its many defects.
The stability landscape features a valley, a lake and a ridge (as described in \cite{Virot2017} and Fig. \ref{fig:stability_landscapes}). Advancing the poker beyond the ridge, defined by the peak poker force, will ultimately bring the system to the lake shore, where the shell buckles. As $F_A$ is increased, the ridge descents, asymptoting towards a zero maximal poker force where the axial load for spontaneous buckling is reached. It is therefore tempting to track the path of the ridge and attempt to infer the critical buckling load without reaching it. Implementing ridge tracking as a non-destructive method to predict the load-carrying capacity of a given shell and rethinking buckling as a nonlinear finite-amplitude instability would enable to finally move beyond the very conservative NASA design rule and develop a fundamental and scalable approach to confront imperfection sensitivity in the design and control of thin shells.

Drawing from the analogy between imperfect shells which buckle and imperfect pipe flow which become turbulent, we show that lateral probing of cylindrical shells can indeed reveal their strength. By examining the safe regions of the stability landscape and extrapolating to the critical conditions, we use lateral probing to non-destructively predict the strength of individual imperfect shells. We direct buckling nucleation, by intentionally introducing a guiding defect in the shell. Doing so, we show, as a proof of concept, that probing near this guiding defect yields an accurate estimate of the shell's buckling load. Our results show that when the guiding defect is identified, tracking the ridge and extrapolating to where it vanishes works extremely well and predicts the critical buckling load to within $5\%$ precision. 
This result considerably surpasses the NASA design rule which estimates a lower bound for the distribution of nominally identical shells while we accurately predict the strength of individual shells. Ridge-tracking thereby allows for a deterministic rather than a probabilistic prediction, and thereby opens avenues towards new design and control strategies for thin shell structures.

\begin{figure*}
\centering
\includegraphics[width = 0.9\textwidth]{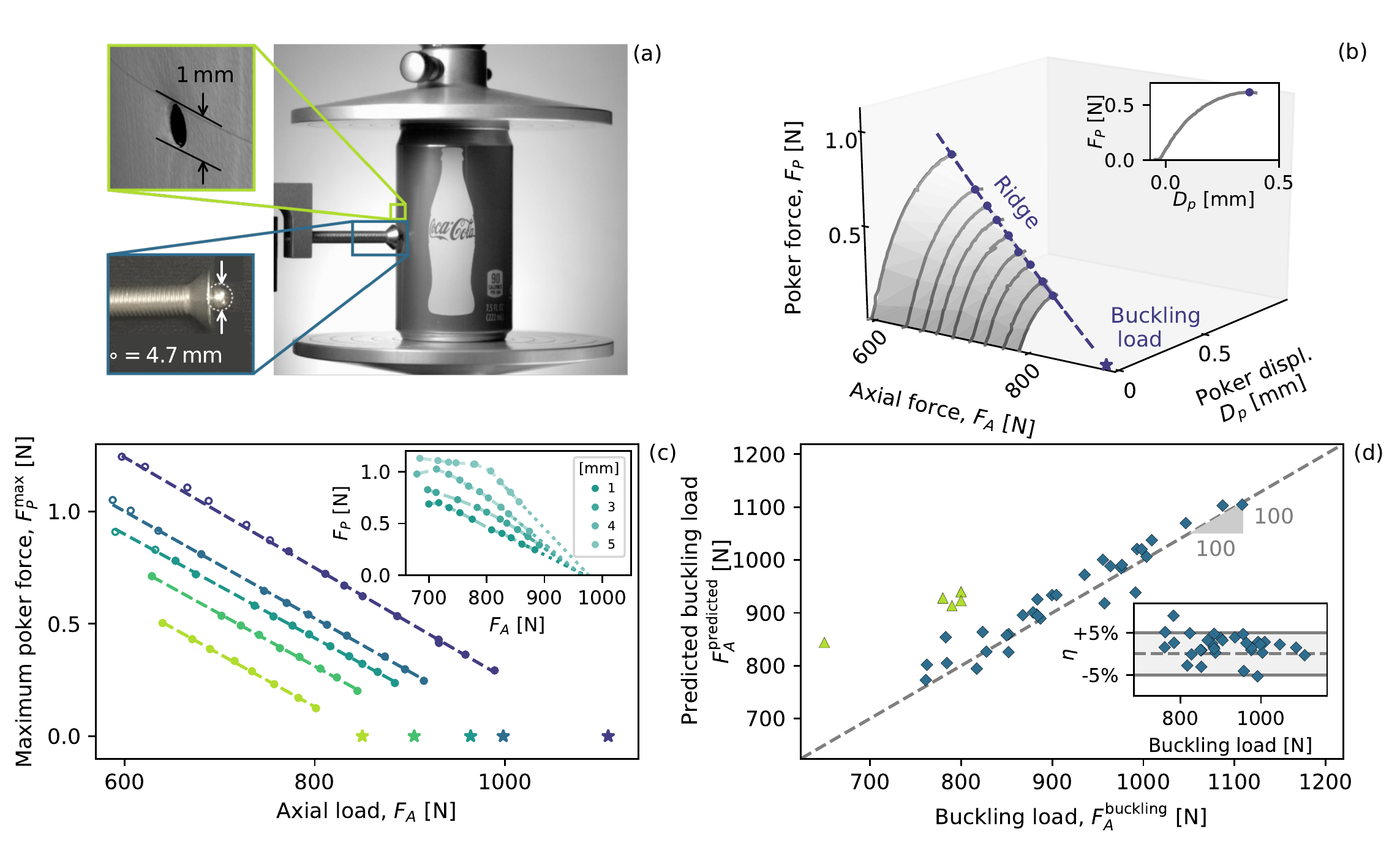}
\caption{
(a) Experimental setup with close-ups of hole and poker tip.
(b) A typical example of ridge tracking. The star indicates the prediction for the spontaneous buckling load. Inset: $F_p$ vs $D_p$ for $F_A=641\,$N. 
(c) Five examples of successfull ridge tracking. Each spontaneous buckling load is indicated by a star. Inset: Ridge of one can for four different poking locations.
(d) Predicted vs. measured spontaneous buckling load for all 38 tested shells. Green triangles indicate failure initiating far from the guiding hole. Inset: Relative error $\eta$.
}
\label{fig:sketch_and_results}
\end{figure*}

The experimental setup consists of a custom-made biaxial mechanical tester (ADMET, Inc.) described in detail previously \cite{Virot2017}. 
A vertical actuator, equipped with a load cell, applies an axial load $F_A$ on the sample by moving at a constant speed of $5\,$mm/min.
The lateral actuator has a steel marble of diameter $4.7\,$mm at its tip and serves as the poker.
Importantly, axial and lateral loading are always displacement-controlled, and lateral poking is performed at set end-shortening; the axial load varies by less than $2\%$ during poking.       
An acquisition card (National Instrument DAQ USB-6001) records forces and displacements, and controls the actuators through an analogue signal, enabling autonomous ridge tracking.
The samples are empty $7.5\,$oz mini Coke cans, made of aluminum. These are cylindrical shells of radius $R = 28.6\,$mm, mid-plane thickness $t = 104\,\pm 4 \,\mu$m (radius-to-thickness ratio $R/t = 274$) and height $L = 107\,$mm. The buckling load of each shell is dictated by the complex correlation structure of its many defects; it is thus practically impossible to predict where buckling will nucleate. 
We therefore intentionally introduce a guiding defect that fixes the preferential nucleation point to a predefined location close to the probe: A hole of $1\,$mm in diameter, is drilled $1\,$mm above the mid-plane of the cylinder as shown in Fig.~\ref{fig:sketch_and_results}(a). The edge of the hole is circular and slightly curved inwards. At the chosen parameters, with a radius below the characteristic scale of $\sqrt{Rt}=1.7\,$mm the strength of the can is not determined by the hole \cite{starnes70,toda83}. Instead the distribution of spontaneous buckling loads remains broad indicating that the stability properties remain dominated by the unknown defects of each specific can.

Ridge-tracking requires systematically poking the sample under gradually-increasing axial loads without triggering buckling, as shown for a typical experiment in Fig.~\ref{fig:sketch_and_results}(b). 
In this example, the sample is initially loaded to an axial load $F_A^0 = 600\pm1$\,N, which is less than $25\%$ of the strength predicted by linear stability analysis. 
The axial end-shortening is then set, and the poker slowly advances towards the sample at a constant rate of $2\,$mm/min, contacting the shell 1\,mm below the hole. 
Once the poker contacts the shell, the probe force gradually increases. 
When the force reaches its peak $F_p^\text{max}$, the poker automatically stops at a displacement $D_p^\text{max}$, and returns to its initial position, as shown in the inset of Fig.~\ref{fig:sketch_and_results}(b). In this example, the point ($F_A = 592$\,N, $D_p^\text{max} = 0.44$\,mm, $F_p^\text{max}= 0.75$\,N) defines the first point of the ridge, as plotted in Fig.~\ref{fig:sketch_and_results}(b). 
As previously reported, this process is reversible and does not damage the sample \cite{Virot2017}.

The axial load is then sequentially increased in steps of approximately $25\,$N and the next poker force-displacement curve is measured.
These force-displacement curves gradually build-up the sample's stability landscape, which exhibits a distinct ridge. 
The procedure is then stopped when the peak poker force $F_p^\text{max}$ falls below $0.3\,$N. 
A linear extrapolation of the ridge to vanishing poker force yields a prediction for the spontaneous buckling load. 
In this specific experiment, ridge-tracking predicts a buckling load of $F_A^\mathrm{predicted}=893\,$N. To test this prediction, the sample is compressed until it spontaneously buckles at a load of $F_A^\mathrm{buckling} = 887\,$N, remarkably close (within 0.5$\%$) to the predicted value as shown in Fig.~\ref{fig:sketch_and_results}(b). 

In total, 38 samples were tested, with initial loads ranging between $F_A^0 = 522\,$N to  $F_A^0 =  694\,$N, and an average load increment of $23\,$N. One exceptionally fragile sample failed at the first loading, before being probed. Otherwise, samples sustained repeated loading cycles, their ridges follow a straight line in the plane ($F_A$, $F_p^\text{max}$), while converging towards a wide range of buckling loads, as shown for 5 examples in Fig.~\ref{fig:sketch_and_results}(c). 

Intriguingly, the ridges are only straight when probing is done near the hole. Probing further away from it yields ridges that turn towards the buckling-load prediction, as shown in the inset to Fig.~\ref{fig:sketch_and_results}(c). 
When probing beyond a distance of approximately 1\,cm from the hole, the ridges do not point towards the actual buckling load, and the prediction fails---hence the importance of introducing a preferred nucleation location and probing close to it.

Overall, for the 38 samples tested, the ridge-tracking protocol accurately predicts the buckling load over a wide range of critical loads, from $F_A^\mathrm{buckling}=761\,$N to $F_A^\mathrm{buckling}=1108\,$N, as shown in  Fig.~\ref{fig:sketch_and_results}(d). The relative error of the prediction, $\eta = (F_A^\mathrm{predicted} - F_A^\mathrm{buckling}) / F_A^\mathrm{buckling} $ 
has a standard deviation of $6.9\%$ for the whole bunch.
Even more remarkably, when buckling does initiate near the hole, this deviation drops to $2.7\%$ and the average error $\eta$ is $1.6\%$ (blue diamonds in Fig.~\ref{fig:sketch_and_results}(d)). This confirms that ridge tracking works extremely well if we probe in the vicinity of the hole guiding nucleation of buckling. 
Besides providing a promising non-destructive method to probe the load-carrying capacity of a shell, these experiments unambiguously show that the ridge is locally distorted by the defects controlling the load-carrying capacity. We now propose an interpretation of the observed behaviour inspired by the analogy with the onset of turbulence in pipe flow, based on a dynamical-system approach.

The dynamical systems approach suggests an appealing conceptual framing for the cylinder buckling problem, illustrated schematically in Fig.~\ref{fig:stability_landscapes}. 
All deformations of a given cylinder span the high dimensional state-space of the system. Specific deformations including the undeformed cylinder, periodic eigenmodes considered in classical stability analysis \cite{Zoelly1915, Koiter1945} but also any local deformation induced by a poker at a specific location are points in the state-space. Thus, deforming the shell with an advancing poker corresponds to tracing out a continuous path in state-space. When the shell is compressed at a sub-critical axial load, the unbuckled state is linearly stable, but non-linearly unstable. Small deformations will relax back to the unbuckled state while large amplitude deformations may trigger buckling. The state-space region around the unbuckled state in which deformations decay forms the finite basin of attraction of the unbuckled state. It can be figuratively thought as a deep concave bowl, the unbuckled state at its base, and all stable deformations roll towards it.

States outside the basin of attraction will not return to the unbuckled state but lead to a nonlinear instability and trigger buckling. Separating deformations that trigger buckling from those that return to the unbuckled state is the basin boundary, a co-dimension one manifold in state-space defining the nonlinear stability threshold. As the axial load increases, the basin of attraction shrinks, until the basin boundary eventually reaches the unbuckled state, as shown in Fig.~\ref{fig:stability_landscapes}(a). At this load, the basin vanishes, the unbuckled state becomes linearly unstable and spontaneous buckling is triggered by infinitesimal perturbations. For a perfect shell, the basin collapses at the critical load computed by standard linear stability theory. 

For a real shell with imperfections, we expect all state-space structures and specifically the basin boundary to be distorted compared to the perfect shell, as shown on Fig.~\ref{fig:stability_landscapes}(b). Consequently, the basin boundary will reach the unbuckled state at a different load, namely the spontaneous buckling load of the defected shell.

Embedded in the basin boundary are unstable saddle points including localized single dimple deformations \cite{Kreilos2017,Horak2006,Audoly2019,Groh2019a} which define the critical amplitude of a localized perturbation triggering buckling nonlinearly. The advancing poker thus probes a path in state-space that leads to the saddle point within the basin boundary.  
Introducing an additional controlled imperfection, such as slightly curved holes deforms the basin of attraction, and likely moves the corresponding single dimple saddle point closer to the unbuckled state. Thus for the real shell, increasing the axial load will trigger the localized unstable mode at the location of the this defect first, which rationalizes how the hole controls the location where buckling is preferentially nucleated.

We suggest the stability landscape is composed of experimentally-accessible projections of the basin of attraction along one path of deformations in a direction corresponding to the single dimple deformations and at different axial loads. The stability landscape thus indicates how the basin of attraction, measured in the direction of the guiding imperfection shrinks under increasing axial load and at which load it vanishes. Probing the shell at one location, passing the ridge and reaching the lake, thus yields the distance between the unbuckled state and the basin boundary, as shown in Fig.~\ref{fig:stability_landscapes}(c). A convenient feature of ridge tracking, as opposed to lake tracking, is that we are tracking the height of the cliff, and not its width, while both vanish at the critical axial buckling load. This probing protocol keeps the system safely within the basin of attraction. 

The successful prediction of spontaneous buckling loads highlights the utility of transferring nonlinear concepts from the study of turbulence transition in subcritical shear flows to shell buckling. Treating buckling as a nonlinear instability triggered by finite-amplitude perturbations of a subcritically loaded shell rather than as linear instability problem, rationalizes the striking sensitivity to defects. Moreover, it provides avenues for predicting the strength of a shell in the presence of an unknown defect distribution.
We have specifically shown that characterizing the load-dependent critical perturbation amplitude via ridge tracking can predict the strength of commercial mini-Coke cans. The method involves introducing an additional localized imperfection in the form of a hole and probing in the vicinity of it. Notably, introducing the hole does not determine the strength of the shell which still varies greatly. In fact the hole does not appear to weaken the can at all; thus, introducing holes does not trivialize the challenge of buckling-load prediction. We hypothesize that the additionally introduced imperfection guarantees that the unstable mode excited first will indeed be a local dimple buckle near the probing location. Under these conditions, the stability landscape provides the most relevant projection of the basin boundary. Without \textit{a priori} knowledge of where buckling will initiate, ridge tracking will likely require to probe at numerous locations. 
The precise conditions under which the additional controlled defect aids probing yet not affects the strength of the shell itself remain to be investigated in detail. Interestingly, while the specific hole does not determine the strength of the can, it does seem to serve as a nucleus or lightning rod for the nucleation of buckling, forcing a specific initiation location and allowing ridge tracking to work. Introducing controlled defects and ridge tracking opens new and exciting approaches for probing and controlling the stability of thin shell structures.

The authors would like to thank John W. Hutchinson and Pedro. M. Reis for insightful discussions. This work was funded by the National Science Foundation through the Harvard Materials Research Science and Engineering
Center DMR-1420570 and the Swiss SNSF under Grant No. 200021-165530. S. M. R. acknowledges support from the Alfred P. Sloan Research Foundation (FG2016-6925).




\begin{thebibliography}{99}

\bibitem{Seide1960} P. Seide, V. I. Weingarten, E. J. Morgan,
The development of design criteria for elastic stability of thin shell structures,
\textit{Space Technology Laboratories}, Report STL/TR-60-0000-19425 (1960).

\bibitem{Yamaki1984}, N. Yamaki, Elastic stability of circular cylindrical shells, \textit{Elsevier} (1984).

\bibitem{NASA_SP8032} V. I. Weingarten, P. Seide, A. L. Braslow,
Buckling of Thin-Walled doubly curved shells,
\textit{NASA Space Vehicle Design Criteria}, NASA SP-8032 (1969).

\bibitem{Weingarten1968} V. I. Weingarten, P. Seide, E. J. Morgan,
Buckling of Thin-Walled Circular Cylinders,
\textit{NASA Space Vehicle Design Criteria}, NASA SP-8007 (1968).

\bibitem{Horton1965} W. H. Horton, S. C. Durham,
Imperfections, a main contributor to scatter in experimental values of buckling load,
\textit{International Journal of Solids and Structures}, 59–72 (1965).

\bibitem{Koiter1945} W. T. Koiter, 
The Stability of Elastic Equilibrium,
\textit{Ph. D. thesis, TH Delft} (1945)

\bibitem{Bushnell981} D. Bushnell,
Buckling of shells, pitfall for designers,
\textit{AIAA Journal} \textbf{19} 1183–1226, (1981).

\bibitem{Davis1987} R. C. Davis, F. Carder,
Buckling tests of a 10-foot diameter stiffened cylinder with rectangular cutouts,
\textit{NASA Technical Memorandum} 88996, (1987).

\bibitem{Singer2002} J. Singer, J. Arbocz, T. Weller,
\textit{Experimental Methods in Buckling of Thin-Walled Structures}, John Wiley \& sons, New York, Volume 1 and 2 (2002).

\bibitem{Hilburger2015} M. W. Hilburger, W. Allen Waters, W. T. Haynie,
Buckling test results from the 8-foot-diameter orthogrid-stiffened cylinder test article TA01,
\textit{NASA Technical Publication} TP-2015-218785, (2015).

\bibitem{Darbyshire1995} A. G. Darbyshire, T. Mullin,
Transition to turbulence in constant-mass-flux pipe flow,
\textit{Journal of Fluid Mechanics}, \textbf{289}, 83-114 (1995).

\bibitem{Grossmann2000} S. Grossmann, 
The onset of shear flow turbulence, \textit{Reviews of Modern Physics}, \textbf{72}, 603-618 (2000).

\bibitem{Kerswell2005} R. Kerswell, 
Recent progress in understanding the transition to turbulence in a pipe, \textit{Nonlinearity}, \textbf{18}, R17-R44 (2005).

\bibitem{Eckhardt2007} B. Eckhardt, T. M. Schneider, B. Hof, J. Westerweel,
Turbulence Transition in Pipe Flow,
\textit{Annual Review of Fluid Mechanics}, \textbf{39}, 447-469 (2007).


\bibitem{Virot2017}
\author{E. Virot, T. Kreilos, T. M. Schneider, S. M. Rubinstein,}
E. Virot, T. Kreilos, T. M. Schneider, S. M. Rubinstein,
Stability landscape of shell buckling,
\textit{Physical Review Letters}, \textbf{119}, 224101 (2017).







\bibitem{Thompson2015} J. M. T. Thompson,
Advances in Shell Buckling: Theory and Experiments,
\textit{International Journal of Bifurcation and Chaos} \textbf{25},  1530001 (2015).
\bibitem{Thompson2016} J. M. T. Thompson, J. Sieber,
Shock-Sensitivity in Shell-Like Structures: With Simulations of Spherical Shell Buckling,
\textit{International Journal of Bifurcation and Chaos} \textbf{26}, 1630003 (2016).
\bibitem{Hutchinson2017a} J. W. Hutchinson, J. M. T. Thompson,
Nonlinear buckling interaction for spherical shells subject to pressure and probing forces,
\textit{Journal of Applied Mechanics} \textbf{84},  061001 (2017).
\bibitem{Hutchinson2017b} J. W. Hutchinson, J. M. T. Thompson,
Imperfections and energy barriers in shell buckling,
\textit{International Journal of Solids and Structures} \textbf{148}, 157-168 (2017).

\bibitem{Groh2019a} Groh R. M. J., Pirrera A.,  
On the role of localizations in buckling of axially compressed cylinders,
\textit{Proc. R. Soc. A} \textbf{475}, 20190006 (2019).
\bibitem{Groh2019b} Groh R. M. J., Pirrera A.,
Spatial chaos as a governing factor for imperfection sensitivity in shell buckling,
\textit{Physical Review E} \textbf{100}, 032205 (2019).
\bibitem{Neville2018} Neville, R. M., Groh R. M. J., Pirrera A., Schenk, M.,
Shape Control for Experimental Continuation,
\textit{Physical Review Letters} \textbf{120}, 254101 (2018).
\bibitem{Lee2016} A. Lee, F. L. Jim{\'e}nez, J. Marthelot, J. W. Hutchinson, and P. M. Reis, 
The geometric role of precisely engineered imperfections on the critical buckling load of spherical elastic shells, 
\textit{Journal of Applied Mechanics}, \textbf{83}, 111005 (2016)
\bibitem{Marthelot2017} J. Marthelot, F. Lopez Jimenez, A. Lee, J. W. Hutchinson, P. M. Reis,
Buckling of a pressurized hemispherical shell subjected to a probing force,
\textit{Journal of Applied Mechanics} \textbf{84}, 121005 (2017).

\bibitem{Kriegesmann2012}  Kriegesmann, B., Hilburger, M., and Rolfes, R., 
The Effects of Geometric and Loading Imperfections on the Response and Lower-Bound Buckling Load of a Compression-Loaded Cylindrical Shell,
\textit{AIAA Paper} \textbf{2012-1864} (2012).
\bibitem{Wagner2018} Wagner, H. N. R., Huhne, C., and Niemann, S.Robust Knockdown Factors for the Design of Spherical Shells Under External Pressure: Development and Validation,
\textit{Int. J. Mech. Sci.} \textbf{141} 58–77 (2018).
\bibitem{Haynie2012} Haynie, W., Hilburger, M., Bogge, M., Maspoli, M., and Kriegesmann, B.,
Validation of Lower-Bound Estimates for Compression-Loaded Cylindrical Shells,
\textit{AIAA Paper} \textbf{2012-1689} (2012).
\bibitem{Arbelo2014} Arbelo, M. A., Degenhardt, R., Castro, S.G.P., Zimmermann, R.,
Numerical characterization of imperfection sensitive composite
structures 
\textit{Composite Structures} \textbf{108} 295-303 (2014)
\bibitem{Hao2015} Hao, P., Wang, B., Tian, K., Du, K., and Zhang, X., 
Influence of Imperfection Distributions for Cylindrical Stiffened Shells With Weld Lands,
\textit{Thin-Walled Struct.} \textbf{93} 177–187 (2015).
\bibitem{Huhne2008} Huhne, C., Rolfes, R., Breitbach, E., and Tessmer, J., 
Robust Design of Composite Cylinder Shells Under Axial Compression—Simulation and Validation,
\textit{Thin-Walled Struct.} \textbf{46} 947–962 (2008).
\bibitem{Jiao2018} P. Jiao, Z. Chen, X. Tang, W. Su, J. Wu,
Design of axially loaded isotropic cylindrical shells using multiple
perturbation load approach – Simulation and validation
\textit{Thin-Walled Struct.} \textbf{133} 1-16 (2018)
\bibitem{Wullschleger2006} Wullschleger, L., 
Numerical Investigation of the Buckling Behaviour o Axially Compressed Circular Cylinders Having Parametric Initial Dimple Imperfections, 
\textit{Doctoral thesis}, ETH, Zurich, Switzerland (2006).
\bibitem{Craig1973} J. I. Craig, M. F. Duggan,
Nondestructive Shell-stability Estimation by a Combined-loading Technique,
\textit{Experimental Mechanics}, \text{September}, 381-388 (1973)


\bibitem{Gerasimidis2018} S. Gerasimidis, E. Virot, J. W. Hutchinson, S. M. Rubinstein,
On Establishing Buckling Knockdowns for Imperfection-Sensitive Shell Structures,
\textit{Journal of Applied Mechanics} \textbf{85}, 091010 (2018).



\bibitem{Zoelly1915} R. Zoelly,
\"Uber ein Knickungsproblem an der Kugelschale, 
\textit{Thesis}, Z\"urich, 19157).

\bibitem{Horak2006} Hor\'ak, J., Lord, G. J., Peletier, M.A., 
Cylinder buckling: the mountain pass as an organizing center,
\textit{SIAM J. Appl. Math.} \textbf{66} 1793–1824 (2006).

\bibitem{Kreilos2017} T. Kreilos, T. M. Schneider,
Fully localized post-buckling states of cylindrical shells under axial compression,
\textit{Proceedings of the Royal Society of London A}, \textbf{473} 20170177 (2017).

\bibitem{Audoly2019} B. Audoly, J. W. Hutchinson,
Localization in Spherical Shell Buckling,
\textit{J. Mech. Phys. Solids} \textbf{136} 103720 (2017).

\bibitem{SIinfo} See supplementary information SI.

\bibitem{toda83}
Toda, S., Buckling of cylinders with cutouts under axial compression. Experimental Mechanics, 23(4), 414-417 (1983).

\bibitem{starnes70}
Starnes, J. H., The effect of a circular hole on the buckling of cylindrical shells (Doctoral dissertation, California Institute of Technology) (1970).
\end{thebibliography}
\end{document}